\documentclass{aa}
\usepackage{natbib}
\usepackage{graphicx}
\usepackage[english]{babel}
\usepackage{txfonts}
\usepackage{hyperref}
\usepackage{tabularx,booktabs}
\usepackage{multirow}
\usepackage{adjustbox}
\usepackage{commath}

\bibpunct{(}{)}{;}{a}{}{,}

\newcommand{\Msun}{$M_{\sun}$}
\newcommand{\Rsun}{$R_{\sun}$}
\begin{document}

\title
{Flares of accretion activity of the 20 Myr old UXOR RZ Psc}

\author
{I.S.\,Potravnov\inst{1},  V.P.\, Grinin \inst{2,3} \and N.A.\,Serebriakova\inst{4}} \institute
{Institute of Solar-Terrestrial Physics, Siberian branch of Russian Academy of Sciences, Lermontov Str. 126A, 664033, Irkutsk, Russia\\
e-mail: ilya.astro@gmail.com
\and Pulkovo Astronomical Observatory, Russian Academy of Sciences, 196140, Pulkovo, St.\,Petersburg, Russia\\
\and V.V.Sobolev Astronomical Institute, Saint-Petersburg State University, Universitetski pr. 28, 198504  St.\,Petersburg,
Russia\\
\and Kazan Federal University, Kremlyovskaya Str. 18, 420008, Kazan, Russia\\}

\date{Received/accepted}

\titlerunning{ }

\authorrunning{Potravnov et al.}

\abstract {We discuss a revision of accretion activity and kinematics of the enigmatic isolated UX Ori type star RZ Psc. Previously, RZ Psc was known to possess only spectroscopic signatures of outflow in the low-excitation lines of alkali metals. The archival high-resolution spectra reveal a short-lived episode of magnetospheric accretion in the system observed via inverse P Cyg profiles at the H$\alpha$ and \ion{Ca}{II} 8542 \AA\ lines. The simultaneous presence of accretion and outflow signatures at \ion{Ca}{II} 8542 \AA\ is suggestive of an accretion-driven origin of the RZ Psc wind. We argue that RZ Psc experiences matter ejection via the magnetic propeller mechanism but variable accretion episodes allow it to sometimes move in the magnetospheric accretion regime. The presence of the weak accretion in the system is also supported by the radiation of the hot accretion spot on the stellar surface observed spectroscopically at the deep photometric minimum of the star. The Galactic motion of RZ Psc calculated with new GAIA DR2 astrometric data suggests possible membership in Cas-Tau OB association with an age of $t=20^{+3}_{-5}$Myr.}
\maketitle

\section{Introduction}

The magnetospheric accretion scenario is now commonly accepted to explain the activity of the young solar-type classical T Tauri stars (CTTS) \citep{Bouvier2007,Hartmann2016}. The accretion from the inner gaseous disc onto a magnetised star is possible if the truncation radius of the stellar magnetosphere $r_{tr}$ is less than the corotation radius $r_{cor}$ in the Keplerian disc. Otherwise, if $r_{tr} > r_{cor}$, then the rotating magnetosphere should expel the accreting matter outwards. This is the so-called magnetic propeller regime originally proposed for the detached binary systems with relativistic accretors \citep{Illarionov1975}. The results of numerical magnetohydrodynamic (MHD) simulations \citep{Romanova2004,Romanova2018} indicated that the propeller mechanism could also be realised in low-mass stars at the pre-main sequence evolutionary stage as confirmed by observations of the magnetic fields and rotational periods of several CTTS \citep{Donati2010,Donati2011,Donati2019}. Recently, we proposed that the propeller mechanism can be responsible for the formation of the narrow blueshifted absorption components (BACs) in \ion{Na}{I} D lines observed in the spectra of some young stars \citep{Grinin2015}. The star RZ Psc is an intriguing object that stimulates this hypothesis and offers further opportunities to investigate.

RZ Piscium (sp:K0 IV) is located at high Galactic latitude ($b \approx -35\degr$) in isolation from the presently active star-forming regions. However, with stellar parameters $T_{ef}=5350\pm150K$, $\lg g=4.2\pm0.2$ dex, $[Fe/H]=-0.3\pm0.05$ dex, and $V\sin i = 12$ km s$^{-1}$ \citep{Potravnov2014}, absorption at \ion{Li}{I} 6708 \AA\ \citep{Grinin2010}, and X-Ray luminosity $\log L_X/L_{bol}\approx -3.2-3.7$ \citep{Punzi2018} (P18) imply that RZ Psc is a young solar-type star. Its UX Ori-type (UXOR) photometric variability \citep{Zaitseva1985,Kiselev1991,Shakhovskoi2003,Kennedy2017} and bright ($L_{IR} / L_{bol} \sim 8\%$) mid-IR ($\lambda \gtrsim 3\mu$m) excess \citep{deWit2013} indicate that the star hosts a circumstellar disc with a small ($\sim 0.4-0.7$ AU) central cavity. This cavity, along with the absence of prominent emission spectrum and $JHK$ IR excess, rarity, and a short duration of the UX Ori-type minima, suggest that RZ Psc could be at a more advanced evolutionary stage than CTTS with a full accretion disc. This inference is further supported by the recent estimation of RZ Psc’s age at $25\pm5$ Myr \citep{Potravnov2013_2}, which exceeds the commonly accepted $\sim$10 Myr timescale of the inner gaseous discs dispersal \citep{Fedele2010}.

Despite its age, RZ Psc demonstrates the spectroscopic footprints of circumstellar activity \citep{Potravnov2013}. The circumstellar gas is manifested in the RZ Psc spectra predominantly in the form of discrete BACs, which are observed in the lines of alkali metals \ion{Na}{I} D, \ion{K}{I} 7699 \AA\, and lines of the \ion{Ca}{II} IR triplet \citep{Potravnov2017}. Such spectroscopic variability could resemble what has been observed in a non-accreting star with a debris disc $\beta$ Pic \citep[][and references therein]{Lagrange1996}. The weak and mostly redshifted absorptions of moderately ionised species such as \ion{Ca}{II} and \ion{Mg}{II},  etc. frequently appear in the $\beta$ Pic spectra on timescales from hours to days. Such variability is thought to be related to the infall and evaporation of star-grazing bodies (exocomets, planetesimals) that cross the line of sight \citep{Ferlet1987,Beust1998}. In contrast, RZ Psc demonstrates exceptionally blueshifted and stronger absorption components presented in its spectra almost permanently with a variation on the timescale of a few days.

We adhere to the explanation that the RZ Psc disc still contains the depleted amount of the primordial gas, which is recycled by the propeller mechanism to the outflow observed in the low-excitation lines of the alkali metals \citep{Grinin2015,Potravnov2017}. The very weak residual emission at H$\alpha$ led us to suspect the existence of accretion in the system and to estimate its upper limit as $\dot{M}$ $\sim$ $7 \cdot 10^{-12}$\Msun yr$^{-1}$ \citep{Potravnov2017}. Unambiguous detection of the accreting material is essential to support our scenario because this gas should also be the raw material for the propeller mechanism.

Recently, P18 published a set of high-quality optical spectra of RZ Psc and noted that some demonstrated possible accretion-related features at H$\alpha$, and we present an in-depth look at this spectroscopic data. We also use GAIA DR2 astrometric data \citep{GAIA1,GAIA2} to investigate the kinematics and origin of RZ Psc as well as to refine its age and parameters.

\section{Observational data}

We used two demonstrative spectra of RZ Psc from the P18 list retrieved from the Keck Observatory Archive. The star was observed with the HIRES spectrograph at the Keck I telescope on 2013 October 21 and November 16 with resolution $R$ = 38 000. One exposure was obtained during the October observations, and three successive exposures separated by $\sim$3$^h$ intervals were obtained on November 16. We also used spectra observed with a 3 m. Shane telescope and Hamilton \'{e}chelle spectrograph ($R$=62000) at the Lick Observatory on the nights of 2013 November 13, 14,  and 2016 August 10. These observations were also published in P18 and shared by C. Melis. Table~\ref{table1} provides a summary on the spectroscopic data. See also Appendix A for observed H$\alpha$ and \ion{Na}{I} 5889 \AA\ profiles.

\begin{figure*}
\begin{center}
\includegraphics[width=0.8\linewidth, angle =0]{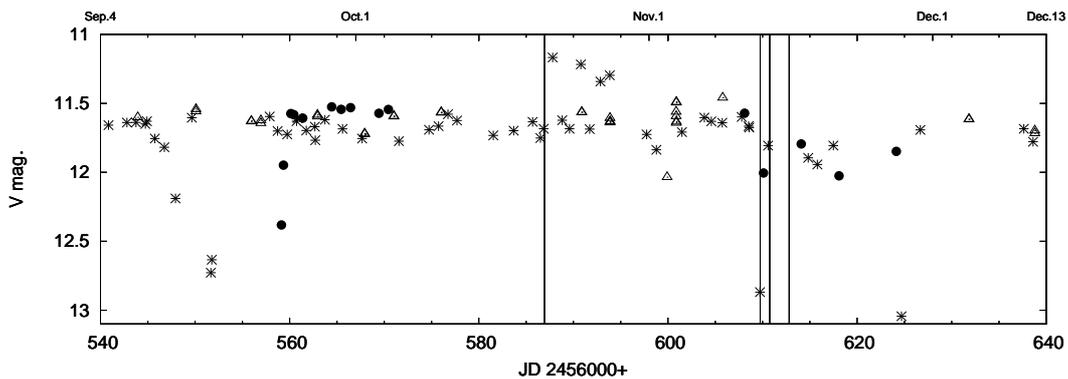}
\caption{\label{1} V band lightcurve of RZ Psc in autumn 2013 compiled from the AAVSO data (asterisks), Sanglok observations (filled circles), and ASAS-SN data (open triangles). Vertical lines mark the date of the spectroscopic observations.}
\end{center}
\end{figure*}

To control the brightness of RZ Psc at the moments of spectroscopic observations, we used the V band photometry from the AAVSO and ASAS-SN \citep{Kochanek2017} databases as well as the observations by N. Minikulov at the Sanglok observatory \citep{Potravnov2014_2}. Fig.~\ref{1} shows the light curve of RZ Psc during the autumn of 2013. According to the AAVSO data on October 21, the star, normally at approximately $V \sim 11^m.5$, had slightly dimmed to $V = 11^m.68$, but on the following night (October 22) RZ Psc had brightened to $V = 11^m.16$. There were no photometric observations of RZ Psc at the night of the observations with HIRES on November 16, but on the following night (November 17) the star was $V = 11^m.8$ in agreement with the observation by N. Minikulov at the Sanglok observatory. Spectroscopic observations on November 13 and 14 caught RZ Psc in a deep photometric minimum and at the end of its egress phase respectively. This result is discussed in detail in Section 4.

\begin{table}[h]
\centering
\caption{Journal of analysed spectroscopic data.}
\label{table1}
\begin{tabular}{l  c c c c}
\hline\hline
\addlinespace
 Date  & Instrument & $\lambda/\Delta\lambda$ & S/N$^1$ & V, mag.\\
\hline
\\
   2013 Oct. 21 & HIRES & 38000 & 115 & 11.68 \\
   2013 Nov. 13 & Hamilton & 62000 & 30 & 12.86\\
   2013 Nov. 14 & Hamilton & 62000 & 35 & 11.8\\
   2013 Nov. 16 & HIRES & 38000 & 130$^2$ & $\sim$11.8\\
   2016 Aug. 10 & Hamilton & 62000 & 90 &11.48\\

\\
  
\hline
\end{tabular}
\tablefoot{$^1$Signal-to-noise ratio (per pixel) was measured for each spectrogram in the region around the H$\alpha$ line. $^2$This value is related to the first (best) among three spectrograms which were obtained on that night. }
\end{table}

\section{Accretion activity of RZ Psc}
\subsection{H$\alpha$ line}
\begin{figure}
\includegraphics[width=\linewidth, angle =0]{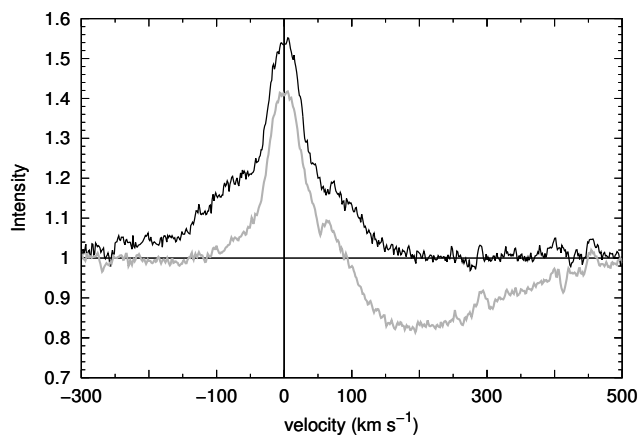}
\caption{\label{2} Residual profiles of H$\alpha$ line on 2013 October 21 (black line) and 2013 November 16 (light grey line). }
\end{figure}

The accretion activity of RZ Psc was evident from the H$\alpha$ profiles observed with HIRES. On the night of October 21, the weak H$\alpha$ emission was seen above the continuum level while the HIRES spectrogram on November 16 revealed the prominent inverse P Cyg type profile (IPC) of the line. Fig.~\ref{2} shows the residuals obtained after subtraction of the photospheric profile of the standard star $\sigma$ Dra from those observed. The star $\sigma$ Dra is the anchor K0 V standard of the MK system with parameters close to those of RZ Psc, $T_{ef}=5270$, $\lg g=4.2$dex, $[Fe/H]=-0.3$dex \citep{Mishenina2013}, and $V\sin i = 1.4$ km s$^{-1}$ \citep{Marsden2014}. The spectrum of the $\sigma$ Dra was retrieved from the ELODIE archive \citep{Moultaka2004} and convolved with corresponding instrumental and rotational kernels to fit the RZ Psc observations.

The residual emission on October 21 shows the composite structure with a relatively narrow ($FWHM$ = 82 km s$^{-1}$) core at stellar velocity superposed on the broad and slightly asymmetric (blueshifted) emission hump. The total equivalent width of this feature is $EW \sim 1.5 \AA\ $. Assuming its origin in the accreting gas during the episode of enhanced accretion in the system, we estimate the mass accretion rate onto RZ Psc as $\dot{M}$ $\sim$ $5 \cdot 10^{-11}$\Msun yr$^{-1}$ following the empirical relation between luminosity in the emission line and full accretion luminosity by \citet{Fang2009}. This new upper limit of accretion rate exceeds the previous value we reported by about an order of magnitude.

The residual H$\alpha$ profile observed on November 16 consists of the zero-velocity emission peak with $FWHM$ = 67 km s$^{-1}$ and redshifted IPC absorption which extension up to $+580$ km s$^{-1}$  was measured in the unsubtracted spectrum. The unsubtracted spectrum was used order to avoid uncertainty due to differences in continuum normalization. Hereafter, velocities are stated in the star's rest frame.


\subsection{Outflow in NaI D lines}

\begin{figure*}
\begin{center}
\includegraphics[width=0.8\linewidth, angle =0]{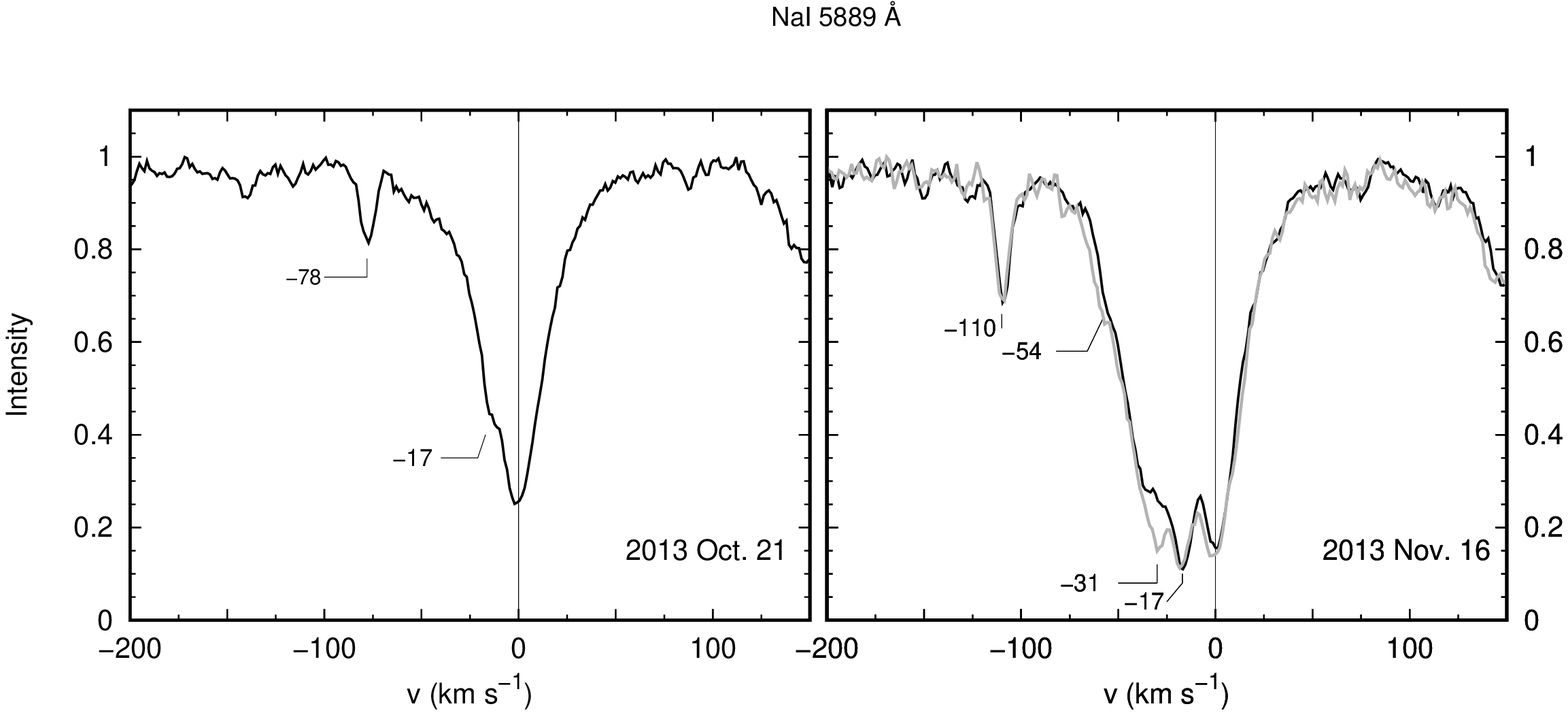}
\caption{\label{3} \ion{Na}{I} 5889 \AA\ line observed in RZ Psc spectra with HIRES. The dates of observation as well as 
the velocities of the BACs (in km s$^{-1}$) are
labeled on the plots. In the plot related to the night of 2013 November 16, the black line shows the profile observed at UT
$4^h 22^m$, and the thick gray line shows the observation at UT  $7^h 30^m$. The \ion{Na}{I} 5895 \AA\ line displays similar behavior.}
\end{center}
\end{figure*}

Simultaneously, the \ion{Na}{I} D lines (Fig.~\ref{3}) show only minor signatures of the wind on October 21 based on the weak BACs at $-78$ km s$^{-1}$ and $-17$ km s$^{-1}$. No signatures of infall are observed by the \ion{Na}{I} D lines on November 16 when H$\alpha$ and \ion{Ca}{II} 8542 \AA\ possessed the IPC profile. The lines of the sodium doublet on that night are disturbed by the deep low-velocity absorptions centred on the velocity around $-30$ km s$^{-1}$ and marginally resolved into three components at its deepest part. The narrow BAC displaced on $-110$ km s$^{-1}$ is observed in clear separation from the low-velocity structure mentioned above. Comparison of the \ion{Na}{I} 5889 \AA\ profiles on the two successive exposures (right panel of Fig.~\ref{3}) reveal the minor growth of the component at $-31$ km s$^{-1}$ within $3^h$. No changes are detected at H$\alpha$ over the same timescale.


\subsection{\ion{Ca}{II} 8542 \AA\ profile}
\begin{figure}
\includegraphics[width=\linewidth, angle =0]{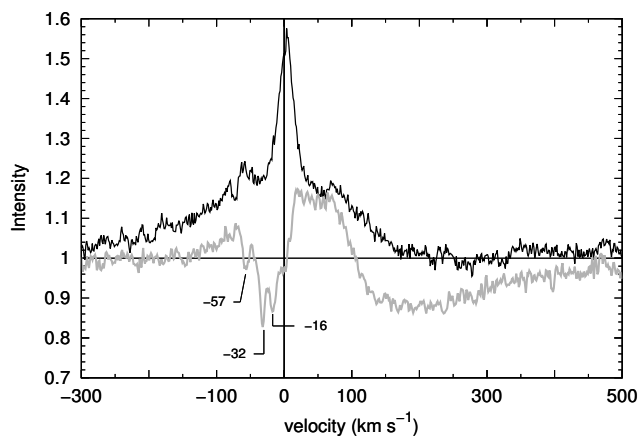}
\caption{\label{4} \ion{Ca}{II} 8542 \AA\ residual profiles in RZ Psc spectra. The observation on 2013 October 21 is shown by the black line, and the 2013 November 16 spectrum is plotted in light gray. The velocities of the BACs in km s$^{-1}$ are labeled on the plots.}
\end{figure}

The \ion{Ca}{II} 8542 \AA\ line displays a most complicated
structure. On the night of October 21, the line profile strongly resembles that of the H$\alpha$ with the same asymmetry of the broad emission hump but with a narrower ($FWHM$ = 25 km s$^{-1}$)  central peak. On the night of November 16, the residual profile appears as the superposition of H$\alpha$ and \ion{Na}{I} D profiles (Fig.~\ref{4}) and simultaneously indicates the wind and the matter infall. This residual consists of the IPC feature similar to H$\alpha$ superposed with the system of BACs at the same velocities as the \ion{Na}{I} D lines. The only exception is that the high-velocity BAC at $-110$ km s$^{-1}$ is absent at \ion{Ca}{II} 8542 \AA\,.


\subsection{Summary on the circumstellar lines in the RZ Psc spectra}

The emission in the H$\alpha$ and \ion{Ca}{II} 8542 \AA\ lines is formed by heated gas as calcium should already be singly ionised with ionisation energy of 6.11 eV, and the formation of the H$\alpha$ emission requires a gas temperature of about $8000K$. Broad components of these lines observed on October 21 possibly originate in the zone of contact between the rotating inner gaseous disc and stellar magnetosphere where gas is shock-heated. The strong ionisation of sodium in this region, as evidenced by the absence of simultaneously observed accretion features in the \ion{Na}{I} D lines, confirms this hypothesis.

According to \citet{Grinin2015} and \citet{Shulman2017}, the \ion{Na}{I} D BACs in the RZ Psc spectra trace the cool part of its wind in the form of a biconical spiral stream. Due to its high spatial porosity, this gaseous stream fills only a minor portion of the complete solid angle $4\pi$ that prevents the formation of substantial emission. For this reason, we do not observe the P Cygni profile in the sodium lines. The low-velocity BACs are formed closer to the star, and the rapid changes on the timescale of $\sim3^h$ reflect the non-stationary processes at the initial stages of the wind acceleration. The long-lived \citep[up to several days, as was shown in][]{Potravnov2017} BAC at $-110$ km s$^{-1}$ in contrast forms at distances of a few dozen stellar radii. This explains its absence in the November 16 spectrum at \ion{Ca}{II} 8542 \AA\ lines due to the higher excitation potential of \ion{Ca}{II} in comparison with \ion{Na}{I} D.

\subsection{Photospheric lines}

To investigate the subtle effects in the photospheric lines, the least-squares deconvolution (LSD) profiles were constructed with a modified (by N. Serebriakova and V. Tsymbal) multicomponent LSD program package \citep{Tkachenko2013}. The differences between the synthetic spectrum with the RZ Psc parameters and its LSD model arising from the principal limitations of the LSD method were calculated. Correction for these differences was applied to the final LSD spectrum constructed using real observations.  
We used the line list from the VALD database \citep{Piskunov1995,Kupka2000,Ryabchikova2015} for the LSD mask and synthetic spectra calculations with SynthV code \citep{Tsymbal1996}. The 5300-5720\AA\ wavelength region was chosen because it is relatively free from the severe line blending and close to the region of the local maximum of optical veiling \citep{Stempels2003,Dodin2012}.
LSD profiles were constructed for three groups of lines, 'weak' with the depth $l < 0.4$, 'medium' ($0.4 \leq l \leq 0.7$), and 'strong' ($l > 0.7 $), and each weighted to the mean depth of the group with 0.2, 0.55, and 0.85, respectively. Thus, the resulting LSD profiles represent the mean profile of the lines with the corresponding depth in the observed spectrum. The profiles for the intermediate depths are obtained with interpolation. The number of spectral lines used in the calculations was 985 in the weak group along with 183 and 103 in the medium and strong groups, respectively.
For further comparison, the 2016 August 10 Hamilton spectrum was selected as the template after visual inspection of the available data. This spectrum was obtained on the bright state (AAVSO: $V=11^m.48$) with scarce signatures of circumstellar activity. The H$\alpha$ profile was in absorption with symmetric wings on this date, and the \ion{Na}{I} D lines showed no clear signatures of the BACs.

Fig.~\ref{5} shows the comparison of the LSD profiles in the RZ Psc spectra at different epochs. The profiles corresponding to the 2013 October 21 observation reveal the shallowness of the strong photospheric lines in comparison with the template. This effect can be interpreted as the well-known T Tauri stars veiling in absorption lines by a non-photospheric emission. The source of this emission is believed to be the hot accretion spot on the stellar surface or accretion shock radiated in the continuum and frequencies of the spectral lines \citep{Basri1990, Dodin2012}. The increased veiling level, along with the enhanced H$\alpha$ emission, observed on this date likely indicate an episode of intensified accretion. The HIRES spectrogram taken on 2013 November 16 shows no notable difference in the LSD profiles compared to the template. In contrast, an even more pronounced veiling of the absorption lines is observed during the deep photometric minimum of the RZ Psc (discussed in the next section).

\section{Spectrum of RZ Psc at the deep photometric minimum}

The unique spectrum of RZ Psc was obtained at Lick Observatory on the night of 2013 November 13 (JD2456609.758). P18 noted the enhanced ($EW$ = 2.3\AA\, from our measurement) double-peaked H$\alpha$ emission above the continuum level (see corresponding profile in Fig.~\ref{A.1.}) and splitting of some individual photospheric lines into two components. They discussed these effects as a probable result of spectroscopic binarity of RZ Psc.

On this observation night, RZ Psc dropped into the deep photometric minimum. The star faded down to $V=12^m.86$ according to AAVSO photometry obtained at JD2456609.733. The subsequent spectroscopic observation made at Lick 2013 November 14 (JD2456610.763) showed RZ Psc brightening from $V=12^m.0$ at the epoch JD2456610.106 (Sanglok obs.) to $V=11^m.8$ at JD2456610.591 (AAVSO).
    
The growth of the H$\alpha$ emission in a deep minimum of the UX Ori type stars is well known and caused by the so-called "coronographic effect" \citep{Grinin1994}, which can be explained in terms of increasing contrast between the gaseous region producing circumstellar H$\alpha$ emission and the stellar continuum. In eclipse, the stellar disc is more effectively screened by the opaque cloud than the much-extended emission region that remains mostly unobscured. \citet{Rodgers2002} also found the emission reversals in some low-excitation metallic lines in the spectrum of RR Tau at minimum light, and explained their formation by the same mechanism.

\begin{figure}
\includegraphics[width=\linewidth, angle =0]{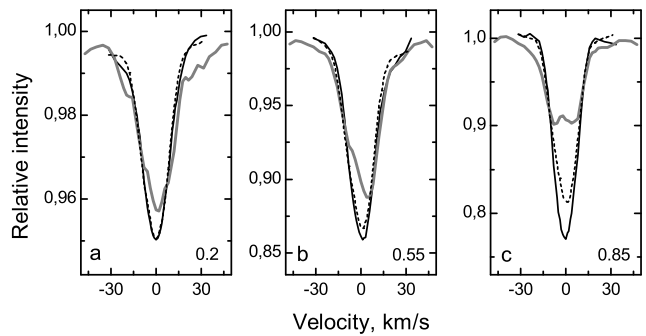}
\caption{\label{5} LSD profiles for 2013 October 21 (dotted line), 2013 November 13 (thick grey line), and 2016 August 10 (thin black line) observations. The three panels represent the (a) weak, (b) medium, and (c) strong lines following the details provided in the text. The scaling factor is labelled in the lower right corner of each panel.}
\end{figure}

The coronographic effect observed at the deep minimum of RZ Psc resulted in the appearance of emission cores in the photospheric lines and impression of their splitting into two distinct components. The LSD profiles (Fig.~\ref{5}) demonstrate that at the eclipse the veiling by emission cores was more prominent in the strongest lines, although it was presented in the entire range of lines depths. The effect persisted in the spectrum obtained during the eclipse egress (2013 November 14), but in reduced form, and was not observed in the template spectrum out of eclipse (2016 August 10).

Such "line-dependent veiling" was detected in some CTTS and attributed to the line emission of a hot accretion spot on the stellar surface \citep{Gahm2008,Dodin2012,Rei2018}. The dependence of the veiling on the line strength is the result of the different depths where the weak and strong photospheric lines are formed. Qualitatively, in the hot spot model, the emission in the spectral line originates due to the temperature inversion, as in solar chromosphere. The stronger lines form in the upper layers of the photosphere more heated by the accretion shock. Thus, the stronger lines are more affected by the veiling effect than the weak ones. The research cited above noted that variability of the veiling factor did not correlate with stellar brightness, and was explained as a result of variability in accretion rates. In the case of RZ Psc, which is a very weak accretor, the effect of screening played a major role in the appearance of emission cores at minimum light. However, the specific geometry of the eclipse is required for such an explanation because the accretion spot should remain unobscured, which can be achieved if the spot is located near the polar regions of the star.

Another important effect revealed with the LSD profiles was the broadening of the absorption lines at eclipse (Fig.~\ref{4}). Such an effect was predicted in the theoretical work by \citet{Grinin2006} and observed in another unusual pre-main-sequence star V718 Per \citep{Grinin2008}. According to their model, the broadening of the photospheric lines in the spectra of young T Tauri stars occurs due to the Doppler scattering of the stellar radiation on the moving dust particles in the inner regions of the circumstellar discs. In eclipse, the contribution of the scattered light to the total stellar radiation increases, and the effect becomes more pronounced. The observed absorption profile becomes the superposition of the photospheric profile broadened by stellar rotation and the profile of the scattered light broadened by the rotation in the inner regions of the circumstellar disc. In eclipse, wings of the RZ Psc LSD profiles could be tentatively fitted by the rotational profile broadened with $V\sin i \approx 21-26.6$ km s$^{-1}$. The out-of-eclipse value of the RZ Psc rotation is $V\sin i = 12$ km s$^{-1}$.
Recent interferometric observations of UXORs confirmed that their discs are seen under large inclination angles of about 70$\degr$ \citep{Kreplin2013,Kreplin2016}. The same inclination is safe to adopt for the RZ Psc disc. Low luminosity of RZ Psc could not produce the puffed-up inner rim in its disc, and the absence of a massive disc wind also eliminates the possibility of raising dust particles far above the disc plane. Thus, the observability of the UX Ori-type photometric minima means the precise tuning of the system is nearly edge-on.

Solving the $\sin i$ uncertainty with $i\sim$70$\degr$ and after profile deconvolution, we obtained the component formed by scattered light was broadened with $V\approx13-23$ km s$^{-1}$. Assuming the Keplerian rotation of the dust, we estimate the radius of the region where the scattered light was produced is 2-5.8 AU from the star.  
 

\section{Kinematics and position on the H-R diagram}

The recently published GAIA DR2 catalogue provides, for the first time, the direct measurement of RZ Psc parallax. According to the catalogue data, the
distance to RZ Psc is $D = 196\pm3.9$ pc and the components of the star's proper motion are $\mu_{\alpha} = 27.37 \pm
0.13$ mas $yr^{-1}$ and $\mu_{\delta} = -12.59 \pm 0.19$ mas $yr^{-1}$. Taking these parameters along with the radial velocity
$V_{R} = -1.2 \pm 0.3$ km s$^{-1}$ \citep{Potravnov2014_2}, we calculated the $UVW$ space velocities of the RZ Psc as $U = -16.9\pm0.5$ km s$^{-1}$, $V = -21.2 \pm0.7$ km s$^{-1}$, and $W = -7.1 \pm0.5$ km s$^{-1}$. Such a combination of the distance and $UVW$ velocities does not allow the attribution of RZ Psc to any known young moving groups (YMG) \citep{Zuckerman2004,Torres2008,Mamajek2016}, including "Pisces MG"\ \citep{Binks2018}, which was recently discovered in the neighbouring region of the sky. Our attempt to identify possibly young co-moving stars with X-ray and IR excesses within 20 pc radius around RZ Psc through the cross-correlation of GAIA DR2, 2RXS \citep{Boller2016}, and ALLWISE \citep{Cutri2013} catalogues was unsuccessful. Previous searches in a 15$\arcmin$ vicinity around RZ Psc by P18 also provided negative results.


\begin{figure*}
\begin{center}
\includegraphics[width=\linewidth, angle =0]{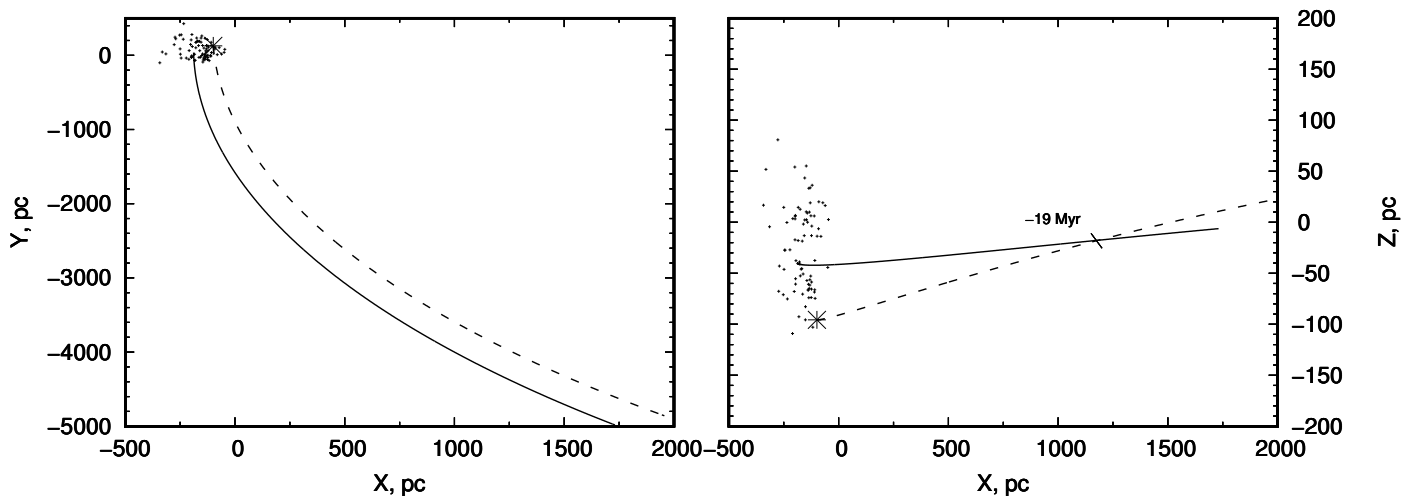}
\caption{\label{6} Galactic orbits of centre of Cas-Tau OB association (solid line) and RZ Psc (dashed line) traced back to 30 Myr and plotted in heliocentric rectangular $XYZ$ coordinates. The intersection point is marked by time in the right panel. The present position of Cas-Tau members from \citet{deZeeuw1999} is shown by small crosses, and the position of RZ Psc is labeled by an asterisk. }
\end{center}
\end{figure*}


However, the GAIA distance and $UVW$ velocities of RZ Psc are close to those of the Cas-Tau OB association \citep{Blaauw1956}. For this old dissolving association, \citet{deZeeuw1999} reported the distance range of 125-300 pc and $(U,V,W) = (-13.24,-19.69,-6.38)$ km s$^{-1}$. They also determined the Galactic coordinates of the convergent point of Cas-Tau as $l_{cp}$ = 243\degr.6, $b_{cp}$ = -13\degr.1, which we used for calculation of the RZ Psc membership probability in association with the modified convergent point method \citep{deBruijne1999}. Assuming the internal velocity dispersion in Cas-Tau as $\sigma_{int} = 3$ km s$^{-1}$, we obtained the membership probability for RZ Psc as $P=52\%$ ($P=77\%$ if we adopt higher internal velocity dispersion $\sigma_{int} = 4.8$ km s$^{-1}$ from \citet{Bhatt2000}). This value is consistent with the cumulative distribution of membership probabilities of the secure members of the Cas-Tau association (see Fig. 30 and Table C1 in \citet{deZeeuw1999}) and, thus, indicate the probable evolutionary relation of RZ Psc to Cas-Tau.

For a detailed investigation of this relation, we calculated the Galactic orbits of the star and centre of Cas-Tau. Orbits were integrated back to -30 Myr in the Miyamoto-Nagai potential \citep{Miyamoto1975} with the \textit{GalPy} numerical integrator \citep{Bovy2015}. In the absence of pronounced spatial concentration of a Cas-Tau association, we used the geometrical centre of the box defining its borders \citep{deZeeuw1999} and adopted the mean distance $D$=200 pc. The resulting calculations are presented in Fig.~\ref{6}. The nearly parallel motion of the RZ Psc and Cas-Tau centres in the Galactic ($XY$) plane are seen while the difference in $W$ velocities leads to the orbits’ divergence in the meridional ($XZ$) plane. The intersection of orbits in the meridional plane occurred at about 20 pc to the south of the Galactic equator around 18-21 Myr ago, taking into account uncertainty in the $W$ velocity of RZ Psc.

\begin{figure}
\includegraphics[width=\linewidth, angle =0]{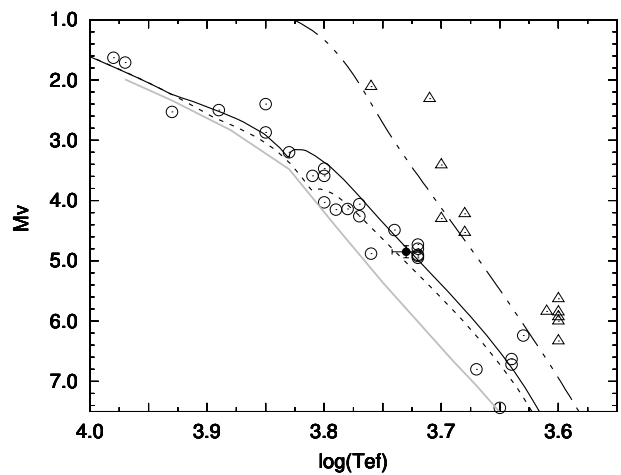}
\caption{\label{7} RZ Psc (black dot) on the H-R diagram. The A-F-G stars of the $\beta$
Pic moving group from \citet{Mamajek2014} and its K-type members from \citet{Shkolnik2017} with GAIA parallaxes are
shown by open circles. Open triangles correspond to the CTTS in Taurus from \citet{Grankin2016}. The 3
(dash-double dot line), 15 (solid line). and 20 (dotted line) Myr isochrones by \citet{Bressan2012} are plotted in the
figure. ZAMS is shown by a thick grey line.}
\end{figure}

With the precise GAIA parallax, we refined the position of RZ Psc on the H-R diagram and specified stellar parameters from the
evolutionary tracks. We calculated the absolute magnitude of the star $M_V = 4^{m}.85\pm0.15$ using its bright state visual magnitude $V = 11^{m}.5$ and the distance $D = 196\pm3.9$ pc. The extinction $A_V = 0^{m}.186$ was determined from the standard extinction law $R_V = 3.1$ and interstellar reddening $E(B-V) = 0.06^{+0.02}_{-0.04}$ per 200 pc given by the 3D map from \citet{Green2018} in the direction toward RZ Psc. On the Fig.~\ref{7}, RZ Psc is placed on the "$M_{V}$ vs $T_{ef}$"\ diagram with isochrones and the Zero Age Main Sequence (ZAMS) from the PARSEC models \citep{Bressan2012} overlaid. For RZ Psc, we used the temperature $T_{ef}$ = 5350$\pm$150K \citep{Potravnov2014}, which coincides within the errors of the recent GAIA value $T_{ef}$ = 5257K. The position of ZAMS was determined from the points of final luminosity and radius minimum on evolutionary tracks at the end of the contraction phase. The corresponding luminosities were converted to observational values $M_V$ with interpolated bolometric corrections from \citet{Kenyon1995}.

From the H-R diagram, RZ Psc lies well above ZAMS between 15-20 Myr isochrones and merges
with the K-type stars of the $\beta$ Pic moving group with the mean age $t=23\pm3$ Myr \citep{Mamajek2014}.
The mass and radius of RZ Psc are $M = 1.1 \pm 0.1$\Msun\ and $R = 1.2 \pm 0.1$\Rsun\ from this position.


\section{Discussion}
\subsection{Accretion activity}

The IPC profile observed on 2013 November 16 in the RZ Psc spectrum at the H$\alpha$ line is a key observational signature of the mass infall predicted by the magnetospheric accretion model. Indeed, such profiles are often observed in the strong permitted lines in the spectra of the accreting CTTS, although with less frequent appearances at H$\alpha$ than in the upper Balmer lines or \ion{Na}{I} D doublet \citep{Edwards1994}. Modelling shows that formation of the IPC feature depends on the optical depth in the considered spectral line as well as on the orientation of the inflow with respect to the observer \citep{Hartmann1994,Muzerolle2001}. The low accretion rate and high inclination of the circumstellar disc to the line of
sight are favourable for the formation of the IPC profile at H$\alpha$. This scenario is exactly what we have in the 
case of RZ Psc where the star is a very weak accretor, and we suppose its disc has an inclination of about $70^{\degr}$.
The red wing of the IPC absorption in H$\alpha$ and \ion{Ca}{II}
8542 \AA\ lines were traced on the unsubtracted November 16 spectrum up to $+580$ km s$^{-1}$. This value exceeds the maximal velocities of about $+200-300$ km s$^{-1}$, which are typically observed in the IPC profiles of the CTTS, and corresponds to the free-fall of the gas accelerated from a distance $\sim 10R_*$. This distance is close to the corotation radius in RZ Psc disc with $$r_{cor} = (GM_*/\omega^2)^{1/3} \approx 9.2R_* \approx0.06AU$$, which was calculated with the updated values of the stellar mass and radius (Section 4), and the projected rotational velocity $V\sin(i) = 12$ km s$^{-1}$. Then, the question remains of how this accretion episode is consistent with the magnetic propeller scenario initially proposed for the explanation of RZ Psc spectroscopic variability.


\begin{figure}
\includegraphics[width=\linewidth, angle =0]{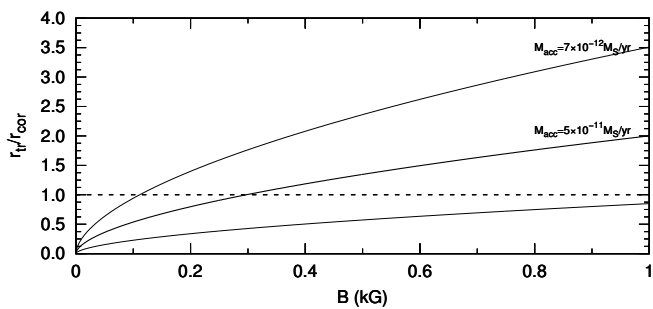}
\caption{\label{8} The dependence of $r_{tr}/r_{cor}$ from the magnetic field strength in the case of RZ Psc. The dotted line indicates the border between the magnetospheric accretion (below) and magnetic propeller (above) regimes. The unlabeled curve was computed assuming a mass accretion rate of $10^{-9}$\Msun yr$^{-1}$ and illustrates pure magnetospheric accretion.}
\end{figure}


The regime of star-disc interaction depends on the interplay between the corotation and truncation radii. The truncation radius is determined by stellar parameters, the mass accretion rate, and the longitudinal component of the global magnetic field expressed as

$$r_{tr}/R_* = 7.1B^{4/7}\dot{M}^{-2/7}_{-8}M_{0.5}^{-1/7}R_2^{5/7}$$, 

where $B$ is the stellar magnetic field strength in kG. $\dot{M}$, $M_{0.5}$, and $R_2$ are the mass accretion rate, stellar mass, and stellar radius, expressed in units of $10^{-8}$\Msun yr$^{-1}$, 0.5\Msun\, and 2\Rsun\, respectively \citep{Bouvier2007}. With available estimations of the mass accretion rate onto RZ Psc of $\dot{M}$ $\sim$ $7 \cdot 10^{-12}$\Msun yr$^{-1}$ \citep{Potravnov2017} and $\dot{M}$ $\sim$ $5 \cdot 10^{-11}$\Msun yr$^{-1}$ (this work), we plot the ratio $r_{tr}/r_{cor}$ versus the strength of the dipolar component of the stellar magnetic field $B$, which is unknown for RZ Psc (Fig.~\ref{8}).

As shown by \citet{Gregory2012}, the strength and topology of the global magnetic fields of pre-main-sequence stars depend on their parameters and internal structure, meaning the position on the H-R diagram. RZ Psc lies in the region on the diagram where the stars held substantial radiative cores. The development of a radiative zone with mass $M_{core}/M_* > 0.4$ is expected to lead to the decrease of the dipole component of the magnetic field that controls the accretion from the disc up to several tenths of kG. From Fig.~\ref{8}, in the case of RZ Psc, such strength of the magnetic field is enough to balance the magnetospheric accretion and magnetic propeller regimes with dependence on the variable mass accretion rate.     

Under the low accretion rate, the stellar magnetosphere expands and can exceed the corotation radius. The magnetocentrifugal barrier blocks the accretion and disperses the gas inflowing from the inner regions of the disc into the surrounding
space. However, the numerical MHD simulation shows that in the case of the weak propeller (applicable to the T Tauri stars), both the accretion and outflow processes are presented \citep{Romanova2018}. These processes are strongly non-stationary and occur in a series of brief episodes. The
accreting matter accumulates in the inner disc, then partially penetrates the magnetocentrifugal barrier infalls
onto the star while the remaining bulk of the gas is ejected as biconical wind through the rotating magnetosphere. Thus, both mechanisms, the magnetospheric accretion and ejection in the propeller regime, occur during such episodes as illustrated by the 2013 November 16 RZ Psc spectrum when we simultaneously observed the accretion at H$\alpha$ via the IPC line profile and intensification of the outflow in the \ion{Na}{I} D and \ion{Ca}{II} 8542 \AA\ lines. However, for the remainder of the time, wind dominates as seen from the simulations mentioned above and observational statistics for RZ Psc.

\subsection{The origin and age of RZ Psc}

The kinematics and distance to RZ Psc obtained with the new GAIA data indicate its probable relation to the Cas-Tau OB association. The traced back Galactic orbits confirm the parallel motion of the star and association projected onto the Galactic plane, while the vertical motion of RZ Psc shows the deviation from the orbit of the poorly defined Cas-Tau centre. Nevertheless, the present position of RZ Psc is consistent with those of the association's members spread significantly over the $Z$ coordinate. Since the moment of divergence in the meridional plane (18-21 Myr ago) is consistent with the isochronal age of RZ Psc (15-23 Myr), assuming it as the moment of RZ Psc’s birth is reasonable. By combining these values, we estimate the median age of RZ Psc as $t=20^{+3}_{-5}$ Myr. The age of the Cas-Tau association is quite uncertain and varies from the isochronal value of 20-30 Myr \citep{deZeeuw1985} to the kinematical age of $\sim50$ Myr \citep{Blaauw1956,deZeeuw1999}. However, both of these determinations (as well as the convergent point coordinates) are based exclusively on early-type members of Cas-Tau while its low-mass stellar content remains unrevealed.  The complete membership census obtained with the advantage of the new GAIA data could refine the kinematical parameters and age of Cas-Tau to more accurately place RZ Psc into its star formation history.

Overall, the refined RZ Psc age is twice as large than the 'classical' 10 Myr milestone of the accretion
termination and inner gaseous disc dispersal \citep{Fedele2010,Williams2011}. However, within the given uncertainty, the
age of RZ Psc can be at least 15 Myr. Recently, the low but statistically meaningful fraction of the
long-lived accretors was found in subgroups of the Scorpius-Centaurus association within a 10-16 Myr range
\citep{Pecaut2016}, and some individual late-type stars possess the accretion signatures at ages up to a few dozen Myr
\citep[see Table 2 in][]{Zuckerman2015,Mamajek2002,Murphy2018}. In light of these results, we suggest that RZ Psc evolved outside of the dense
cluster environment preserves its protoplanetary disc until the present time. Thus, the primordial origin of gas and dust in RZ Psc system in our opinion is a more natural explanation than collisional replenishment of the circumstellar material due to planetesimals grinding \citep{deWit2013} or even destruction of the giant planet (P18).

\section{Conclusions}

The high-resolution HIRES spectra revealed that RZ Psc occasionally demonstrates short-living accretion events consistent with the magnetospheric accretion scenario. The IPC profile at the H$\alpha$ line observed on 2013 November 16 indicates matter infall onto the star from a distance of about $10R_*$, which matches with the corotation radius in the RZ Psc disc. The \ion{Ca}{II} 8542 \AA\ profile simultaneously demonstrates the BACs similar to those observed in the \ion{Na}{I} D lines and IPC profile. This complex structure of the \ion{Ca}{II} 8542 \AA\ profile indicates unequivocally the accretion-related nature of the cool anisotropic wind traced in the RZ Psc spectra by discrete structures in the low-excitation lines of alkali metals. The observational statistics now supplemented by the infall event indicate that both the magnetospheric accretion and propeller ejection are possible in the RZ Psc system. On the short timescale the regime of interaction between the stellar magnetosphere and depleted inner gaseous disc depends on the variable accretion rate, which according to our estimations, varies in the range $\dot{M}$ $\sim$ $7 \cdot 10^{-12} - 5 \cdot 10^{-11} $\Msun yr$^{-1}$.    

The kinematics of RZ Psc obtained with the new GAIA DR2 data revealed the probable origin of the star in the Cas-Tau OB association of about $t=20^{+3}_{-5}$Myr ago. This age is consistent with the very low accretion rate in the system and indicates that RZ Psc is the extreme but not exceptional case of prolonged primordial disc evolution. Thus, the star is a marker object for a detailed study of how the accretion decays and which mechanisms play a role during the later stages of accretion activity. For a deeper understanding of these processes, measurements of the RZ Psc magnetic field are desirable. Spectroscopic monitoring with better temporal sampling over a few stellar rotational cycles also can be useful to clarify the connection between the accretion events and launching of the wind.

\begin{acknowledgements}
We are grateful to Carl Melis for sharing the Hamilton spectra of RZ Psc and providing useful comments on the manuscript. This research was supported by the grant RFBR 18-32-00501. The analysis of the photospheric lines profiles was made with the support of budgetary funding from the Basic Research program II.16.
We also acknowledge with thanks the Keck Observatory Archive (KOA), which is operated by the W. M. Keck Observatory and the NASA Exoplanet Science Institute (NExScI), under contract with the National Aeronautics and Space Administration and the variable star observations from the AAVSO International Database contributed by observers worldwide and used in this research.
We thank the referee for comprehensive and constructive reviewing of the manuscript which helped to improve it.
\end{acknowledgements}

\bibliographystyle{aa.bst}
\bibliography{reference.bib}

\begin{appendix}
\section{Observed H$\alpha$ and \ion{Na}{I} 5889 \AA\ profiles}
\nopagebreak
\begin{figure*}
\begin{center}
\includegraphics[width=0.8\linewidth, angle =0]{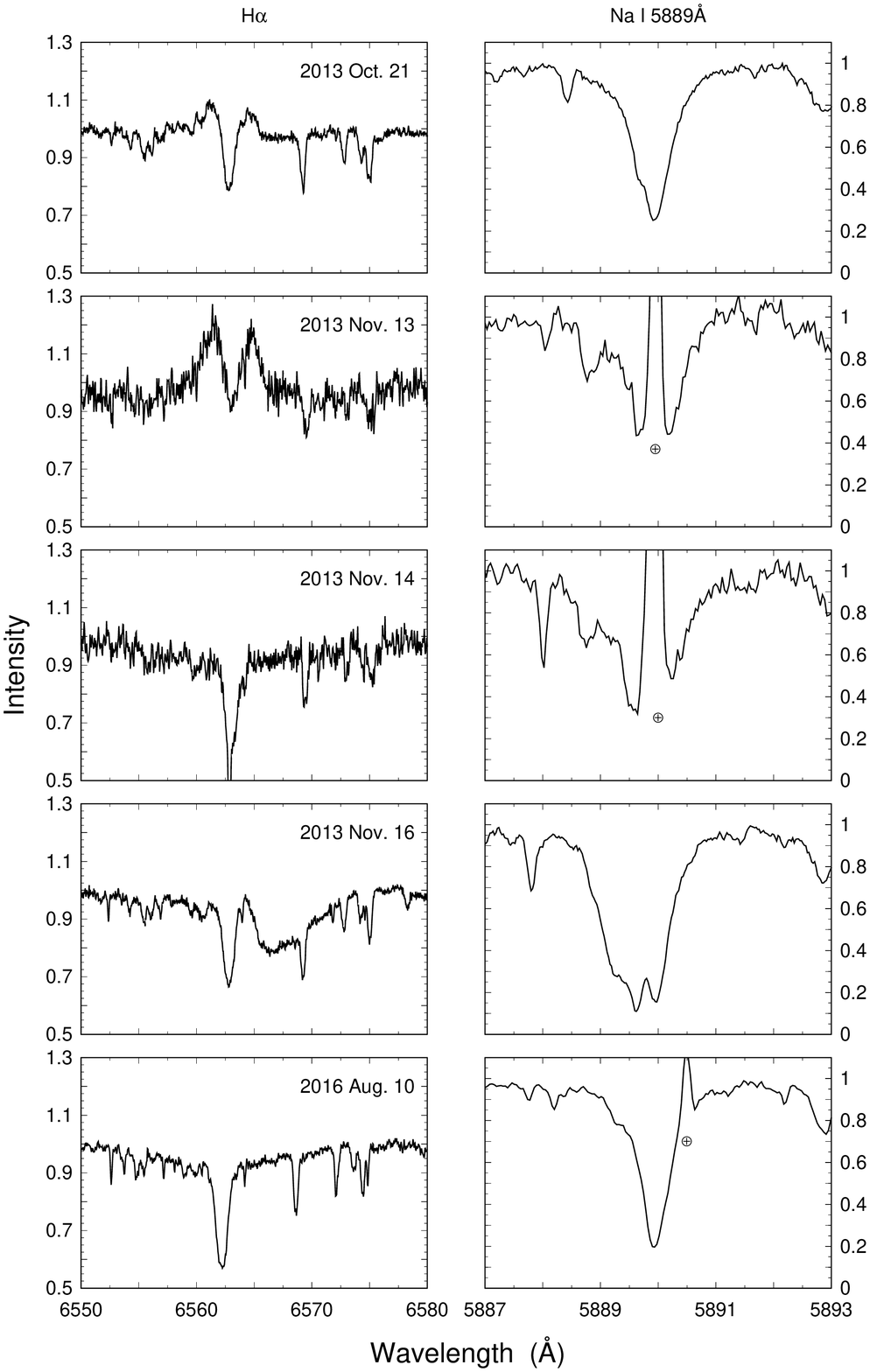}
\caption{\label{A.1.} H$\alpha$ and \ion{Na}{I} 5889 \AA\ profiles observed in RZ Psc spectra with HIRES and Hamilton spectrographs (see Table~\ref{table1} for observational log). 
Sodium lines observed with Hamilton (2013 November 13 and 14, 2016 August 10) are contaminated by the terrestrial "city light" emissions which are marked by the Earth symbol. }
\end{center}
\end{figure*}


\end{appendix}

\end{document}